\documentclass[english,prl,reprint,superscriptaddress,nolongbibliography,twocolumn]{revtex4-1}
\usepackage[T1]{fontenc}
\usepackage[latin9]{inputenc}
\usepackage{verbatim}
\usepackage{textcomp}
\usepackage{amsmath}
\usepackage{amssymb}
\usepackage{graphicx}

\makeatletter

\providecommand{\tabularnewline}{\\}

\@ifundefined{textcolor}{}
{%
 \definecolor{BLACK}{gray}{0}
 \definecolor{WHITE}{gray}{1}
 \definecolor{RED}{rgb}{1,0,0}
 \definecolor{GREEN}{rgb}{0,1,0}
 \definecolor{BLUE}{rgb}{0,0,1}
 \definecolor{CYAN}{cmyk}{1,0,0,0}
 \definecolor{MAGENTA}{cmyk}{0,1,0,0}
 \definecolor{YELLOW}{cmyk}{0,0,1,0}
}

\usepackage{graphicx}

\usepackage{bibunits}
\defaultbibliography{citations}
\defaultbibliographystyle{apsrev}

\makeatother

\usepackage{babel}
\begin{document}
\begin{bibunit}

\title{Mixed Dimensionality of Confined Conducting Electrons in the Surface
Region of SrTiO$_{3}$}

\author{N.~C.~Plumb}

\affiliation{Swiss Light Source, Paul Scherrer Institut, CH-5232 Villigen PSI,
Switzerland}

\author{\negthickspace{}$^{,*}$ M.~Salluzzo}

\affiliation{CNR-SPIN, Complesso Universitario Monte S. Angelo, Via Cinthia I-80126,
Napoli, Italy}

\author{E.~Razzoli}

\affiliation{Swiss Light Source, Paul Scherrer Institut, CH-5232 Villigen PSI,
Switzerland}

\author{M.~M\aa{}nsson}

\affiliation{Laboratory for Neutron Scattering \& Imaging, Paul Scherrer Institut,
CH-5232 Villigen PSI, Switzerland}

\affiliation{Institute of Condensed Matter Physics, \'{E}cole Polytechnique F\'{e}d\'{e}rale
de Lausanne (EPFL), CH-1015 Lausanne, Switzerland}

\affiliation{Laboratory for Solid State Physics, ETH Z\"{u}rich, CH-8093 Z\"{u}rich,
Switzerland}

\author{M.~Falub}

\author{J.~Krempasky}

\affiliation{Swiss Light Source, Paul Scherrer Institut, CH-5232 Villigen PSI,
Switzerland}

\author{C.~E.~Matt}

\affiliation{Swiss Light Source, Paul Scherrer Institut, CH-5232 Villigen PSI,
Switzerland}

\affiliation{Laboratory for Solid State Physics, ETH Z\"{u}rich, CH-8093 Z\"{u}rich,
Switzerland}

\author{J.~Chang}

\affiliation{Swiss Light Source, Paul Scherrer Institut, CH-5232 Villigen PSI,
Switzerland}

\affiliation{Institute of Condensed Matter Physics, \'{E}cole Polytechnique F\'{e}d\'{e}rale
de Lausanne (EPFL), CH-1015 Lausanne, Switzerland}

\author{M.~Schulte}

\author{J.~Braun}

\author{H.~Ebert}

\affiliation{Department Chemie, Ludwig-Maximilians-Universit\"{a}t M\"{u}nchen,
81377 M\"{u}nchen, Germany}

\author{J.~Min\'{a}r}

\affiliation{Department Chemie, Ludwig-Maximilians-Universit\"{a}t M\"{u}nchen,
81377 M\"{u}nchen, Germany}

\affiliation{New Technologies --- Research Center, University of West Bohemia,
Univerzitni 8, 306 14 Pilsen, Czech Republic}

\author{B.~Delley}

\affiliation{Condensed Matter Theory Group, Paul Scherrer Institut, CH-5232 Villigen
PSI, Switzerland}

\author{K.-J.~Zhou}

\affiliation{Swiss Light Source, Paul Scherrer Institut, CH-5232 Villigen PSI,
Switzerland}

\author{\negthickspace{}$^{,\dagger}$ T.~Schmitt}

\author{M.~Shi}

\affiliation{Swiss Light Source, Paul Scherrer Institut, CH-5232 Villigen PSI,
Switzerland}

\author{J.~Mesot}

\affiliation{Swiss Light Source, Paul Scherrer Institut, CH-5232 Villigen PSI,
Switzerland}

\affiliation{Institute of Condensed Matter Physics, \'{E}cole Polytechnique F\'{e}d\'{e}rale
de Lausanne (EPFL), CH-1015 Lausanne, Switzerland}

\affiliation{Laboratory for Solid State Physics, ETH Z\"{u}rich, CH-8093 Z\"{u}rich,
Switzerland}

\author{L.~Patthey}

\affiliation{Swiss Light Source, Paul Scherrer Institut, CH-5232 Villigen PSI,
Switzerland}

\affiliation{SwissFEL, Paul Scherrer Institut, CH-5232 Villigen PSI, Switzerland}

\author{M.~Radovi\'{c}$^{\ddagger,}$}

\affiliation{Swiss Light Source, Paul Scherrer Institut, CH-5232 Villigen PSI,
Switzerland}

\affiliation{Institute of Condensed Matter Physics, \'{E}cole Polytechnique F\'{e}d\'{e}rale
de Lausanne (EPFL), CH-1015 Lausanne, Switzerland}

\affiliation{SwissFEL, Paul Scherrer Institut, CH-5232 Villigen PSI, Switzerland}
\begin{abstract}
Using angle-resolved photoemission spectroscopy, we show that the
recently-discovered surface state on SrTiO$_{3}$ consists of non-degenerate
$t_{2g}$ states with different dimensional characters. While the
$d_{xy}$ bands have quasi-2D dispersions with weak $k_{z}$ dependence,
the lifted $d_{xz}$/$d_{yz}$ bands show 3D dispersions that differ
significantly from bulk expectations and signal that electrons associated
with those orbitals permeate the near-surface region. Like their more
2D counterparts, the size and character of the $d_{xz}$/$d_{yz}$
Fermi surface components are essentially the same for different sample
preparations. Irradiating SrTiO$_{3}$ in ultrahigh vacuum is one
method observed so far to induce the ``universal'' surface metallic
state. We reveal that during this process, changes in the oxygen valence
band spectral weight that coincide with the emergence of surface conductivity
are disproportionate to any change in the total intensity of the O
$1s$ core level spectrum. This signifies that the formation of the
metallic surface goes beyond a straightforward chemical doping scenario
and occurs in conjunction with profound changes in the initial states
and/or spatial distribution of near-$E_{F}$ electrons in the surface
region.
\end{abstract}

\pacs{73.20.-r, 79.60.-i}

\maketitle
SrTiO$_{3}$ (STO) is a foundational material for the coming age of
multifunctional oxide devices. Perhaps most famously, it hosts quasi-2D
conducting states at interfaces with various transition metal oxides
\cite{Ohtomo2004,Hotta2007,Perna2010,Chen2011b,Chen2013}. %
Moreover, it was recently shown that a low-dimensional metal can form
on the bare surface of STO \cite{Santander-Syro2011,Meevasana2011,DiCapua2012}.
The discovery promises to extend this material's technological importance,
as well as shed new light on the physics of metallic oxide surfaces
and interfaces in general, so long as the properties and origin of
the state can be understood and harnessed.

Although stoichiometric bulk STO is an insulator with a 3.2-eV bandgap,
photoemission experiments have observed metallicity in or on STO for
many years \cite{Courths1989,Aiura2002}. However, the electronic
structure and low-dimensional nature of the metallicity had not been
deeply appreciated until very recent angle-resolved photoemission
spectroscopy (ARPES) and scanning tunneling spectroscopy (STS) studies
\cite{Santander-Syro2011,Meevasana2011,DiCapua2012}. The ARPES measurements
have revealed a highly 2D subband structure whose circular Fermi surface
(FS) components and polarization selection rules indicate $d_{xy}$
symmetry. However, depending on the measurement conditions, the spectra
have occasionally glimpsed shallow bands consistent with $3d_{xz}$/$3d_{yz}$
states coexisting with the $3d_{xy}$ subbands \cite{Meevasana2010,Santander-Syro2011}.
The observations of these additional bands allude to a more complex
FS topology than that of the $d_{xy}$ subbands alone, but the properties
of the shallow bands and their relationship to the surface state has
so far not been deeply studied. Additionally there are still open
questions about the origin of the surface metallicity and the spectroscopic
signatures associated with its formation. %

To investigate these issues, we performed ARPES and core level x-ray
photoemission spectroscopy (XPS) to study STO(001) wafers that were
initially prepared to be highly TiO$_{2}$-terminated \cite{SupMat}%
. Just prior to the photoemission measurements, each sample was annealed
\emph{in situ} at 550 $^{\circ}$C in 100 mbar of O$_{2}$ for about
2 hours in order to establish a nominally oxygen-filled starting point.
Certain samples then underwent subsequent \emph{in situ} UHV annealing
procedures in order to generate oxygen vacancies or other defects,
thereby changing the nominal doping of each sample. In addition, one
sample was lightly Nb-doped (0.25\% by weight, Nb-STO). More details
of the sample treatments can be found in the Supplemental Material
\cite{SupMat}. The resulting surfaces were studied without cleaving.

\begin{figure}[!h]
\includegraphics[width=1\columnwidth]{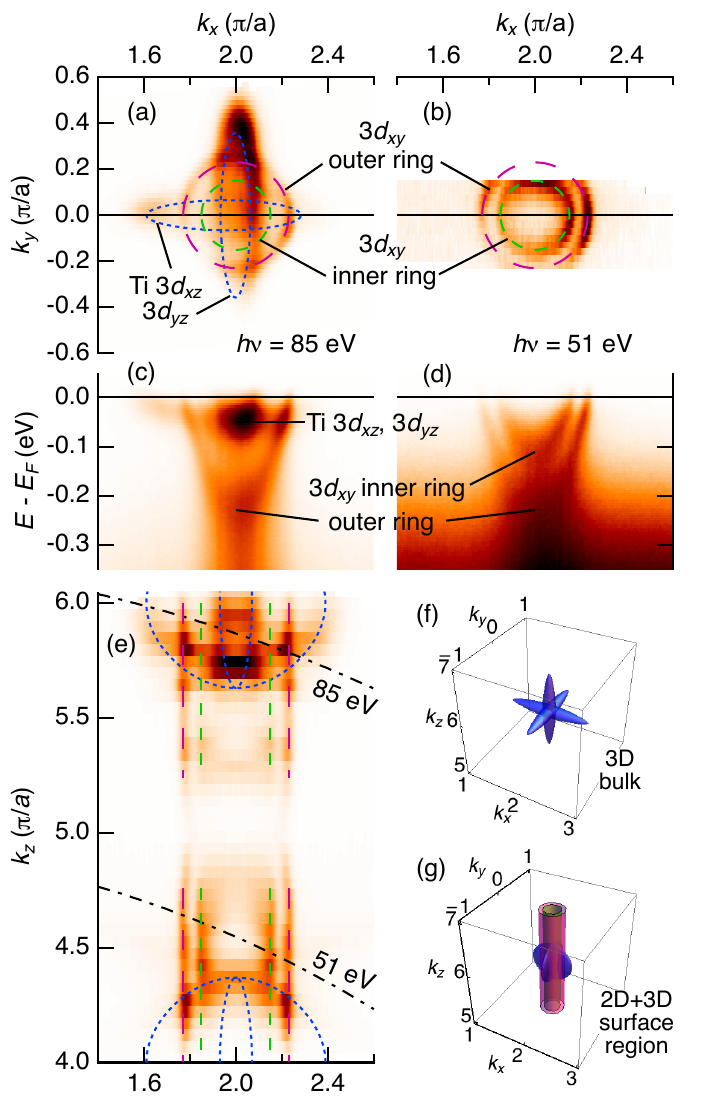}\protect\caption{\label{fig:fig1} (color online.) Three-dimensional view of the near-$E_{F}$
electronic structure of the metallic surface region on STO. (a) Fermi
surface map in the $k_{x}\text{-}k_{y}$ plane measured at $h\nu=85$
eV. The data are from the Brillouin zone centered at $(k_{x},k_{y})=(2\pi/a,0)$.
The ellipsoidal Ti $3d_{xz}$ and $3d_{yz}$ bands are illustrated
by dotted blue lines, while the Ti $3d_{xy}$ inner and outer rings
are highlighted by short-dashed green lines and long-dashed magenta
lines, respectively. (b) Analogous data taken at $h\nu=51$ eV. (c),
(d) Band dispersions along $k_{x}$ at $k_{y}=0$ for each of the
above panels. (e) Fermi surface cut in the $k_{x}\text{-}k_{y}$ plane
at $k_{y}=0$. The dot-dashed lines indicate the curvature of FS cuts
in (a) and (b). (f) Expected shape of the 3D Fermi surface in the
bulk. For reference, the FS volume shown here corresponds to a carrier
density of about $4\times10^{20}$ cm$^{-3}$. (g) Simplified representation
of the mixed quasi-2D and 3D Fermi surface sheets at the STO surface.
The colors correspond with the lines in (a)--(b).}
\end{figure}

The ARPES measurements reveal four FS components, which are highlighted
in Fig.~\ref{fig:fig1}. The data shown come from Nb-STO, but all
the samples are virtually identical in terms of the electronic structure
at the surface. When measured using a photon energy of $h\nu=85$
eV, the FS in the surface $k_{x}\text{-}k_{y}$ plane is made up of
two concentric rings (the inner of which has only very weak intensity),
as well as two ellipsoids aligned along the $k_{x}$ and $k_{y}$
directions {[}Fig.~\ref{fig:fig1}(a){]}. At other photon energies,
the ellipsoids vanish while the rings remain, as demonstrated at $h\nu=51$
eV {[}Fig.~\ref{fig:fig1}(b){]}. Figures \ref{fig:fig1}(c) and
(d) show the corresponding dispersion cuts along $(k_{x},k_{y}=0)$
for (a) and (b), respectively. Based on their $k_{x}\text{-}k_{y}$
symmetry, the ellipsoids can be associated with Ti $3d_{xz}$ / $3d_{yz}$
orbitals and the rings with Ti $3d_{xy}$ orbitals. The inner ring
has been proposed to be a quantum well subband of the outer ring \cite{Santander-Syro2011,Meevasana2011}
(i.e., the $n=2$ quantum well state, while the outer ring is $n=1$)%
{} %
, but new experiments suggest the pair is instead related to Rashba-like
spin-orbit coupling \cite{Santander-Syro-unpublished}. The rings
had already been considered within the context of the surface metallic
state \cite{Santander-Syro2011,Meevasana2011}. Until now, however,
the ellipsoids had never been fully characterized, and there were
conflicting assessments of their dimensionalities and possible relation
to the surface \cite{Meevasana2010,Santander-Syro2011}.

In ARPES, the different photon energies used to probe the FS in the
$k_{x}\text{-}k_{y}$ plane correspond to different planes cutting
across the $k_{z}$ axis \cite{Damascelli2004}. Thus by varying the
photon energy, we mapped the complete FS of the metallic state on
STO in 3D. Figure \ref{fig:fig1}(e) shows the structure of the FS
evaluated as a function of $\boldsymbol{k}=(k_{x},0,k_{z})$. The
outer ring has highly 2D character, though with a slight deviation
in the Fermi momentum $k_{F}$ near the Brillouin zone boundary at
$k_{z}=5$ $\pi/a$. The inner ring is fairly cylindrical with long
parallel segments, but near the zone boundaries along $k_{z}$ it
appears to close to form \textquotedblleft endcaps\textquotedblright{}
of a pill-shaped FS component. These endcaps likely signal a slight
departure from a perfect 2D state, as similarly suggested by the warping
of the outer ring near the zone boundary. We hence regard the inner
and outer ring $d_{xy}$ states as quasi-2D. Otherwise the endcaps
may result from complicating factors such as matrix element effects
or scattered weight due to an out-of-plane reconstruction, although
so far we do not find clear evidence to support these scenarios. 

Most interestingly, the $d_{xz}$ and $d_{yz}$ bands, which in bulk
calculations \cite{Aiura2002,Meevasana2010} are expected to be prolate
spheroids {[}\textquotedblleft cigars\textquotedblright , Fig.~\ref{fig:fig1}(f){]},
are actually stretched along the $k_{z}$ axis {[}\textquotedblleft flying
saucers\textquotedblright , Fig.~\ref{fig:fig1}(g){]}. As a result,
while in the $k_{x}$-$k_{y}$ plane all carriers have effective masses
in line with bulk expectations \cite{Aiura2002}, the strong elongation
of the $d_{xz}$/$d_{yz}$ bands in $k_{z}$ corresponds to a high
effective mass in the out-of-plane direction ($m_{z}^{*}\approx15m_{e}$).
(See Supplemental Material for more details about the extracted effective
masses \cite{SupMat}.) However, the $d_{xz}$/$d_{yz}$ states, while
distinct from truly bulklike electrons due to their substantially
different $z$-axis dispersions, nevertheless show 3D character by
virtue of their fully closed FS components along all $\boldsymbol{k}$
directions. We thus conclude that the $d_{xz}$ and $d_{yz}$ electrons
penetrate multiple unit cells toward the bulk, while the $d_{xy}$
electrons are more tightly confined to the surface. The overall picture
is similar to the predicted orbital-resolved distribution of carriers
in STO near the LaAlO$_{3}$/STO(001) interface \cite{Delugas2011,Zhong2013a}.
The measurements are in good qualitative and quantitative agreement
with previous observations from variously annealed cleaved samples
studied by ARPES using only select photon energies \cite{Meevasana2011,Santander-Syro2011}.
Thus the results here tie prior findings together and account for
the visibility or invisibility of the ellipsoids in previous ARPES
spectra of the metallic surface, which can be attributed the dimensionality
of these states and the choice of measurement conditions (specifically
the photon energy and Brillouin zone) that affect the momentum space
being probed.

\begin{figure}
\includegraphics[width=1\columnwidth]{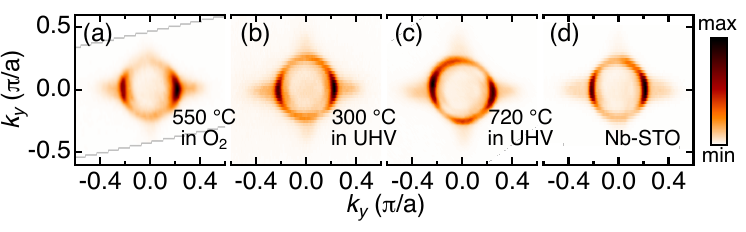}\protect\caption{\label{fig:fig2} (color online.) Universality of the FS with respect
to annealing conditions and light bulk doping. (a)--(d) Fermi surfaces
of various STO samples (annealed in O$_{2}$ at 550 $^{\circ}$C,
annealed in UHV at 300 $^{\circ}$C, annealed in UHV at 720 $^{\circ}$C,
and lightly Nb-doped, respectively). The measurements were performed
in the first Brillouin zone using $h\nu=85$ eV, corresponding to
roughly $k_{z}=6.2\pi/a$, thus intersecting with the $d_{xz}$/$d_{yz}$
ellipsoids.}
\end{figure}

Like the quasi-2D rings \cite{Santander-Syro2011}, the sizes of the
ellipsoids are essentially universal with respect to bulk oxygen vacancies
or dopants (e.g., regardless of whether the samples are transparent
or black), further confirming that these FS components are associated
with the near-surface region, despite their 3D nature. This is illustrated
in Figs.~\ref{fig:fig2}(a)--(d), which show FSs measured on STO
samples prepared by various \emph{in situ} annealing treatments, as
well as bulk Nb-doped STO.

\begin{figure}
\includegraphics[width=1\columnwidth]{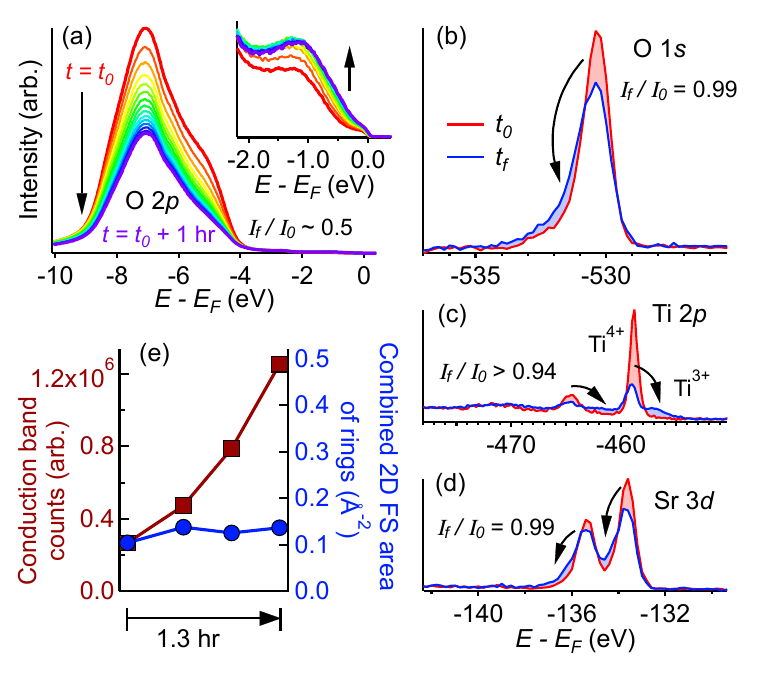}\protect\caption{\label{fig:fig3} (color online.) Spectral evolution of the sample
in Fig.~\ref{fig:fig2}(a) as a function of irradiation time. (a)
Decrease of the O $2p$ valence band spectral intensity during irradiation.
The inset highlights in-gap and metallic states that form during the
same period. (b) XPS spectra of the O $1s$ core level measured before
($t_{0}$) and after ($t_{f}$) a similar radiation dose as in (a).
(c), (d) Analogous spectra for the Ti $2p$ and Sr $3d$ core levels.
(e) Comparison of the time evolution of total integrated counts in
the conduction band region and the combined $k$-space areas of the
inner and outer $3d_{xy}$ ring-shaped FS components.}
\end{figure}

So far it appears there may be multiple methods for preparing the
metallic surface state on STO \cite{Santander-Syro2011,DiCapua2012},
including exposing the material to synchrotron radiation under UHV
conditions ($\sim10^{-11}$ mbar) typical for ARPES \cite{Meevasana2011}.
In Fig.~\ref{fig:fig3}(a), starting from an insulating, oxygen-annealed
sample {[}the same as in Fig.~\ref{fig:fig2}(a){]} that initially
shows no FS, we expose a previously unstudied spot on the sample to
the beam for an initial time $t_{0}$ ($\sim10$ minutes using $h\nu=47$
eV) to establish the onset of surface metallicity. Sample charging
is alleviated by a grounding technique described in the Supplemental
Material \cite{SupMat}. During the beam exposure, the spectral weight
associated with the O $2p$ valence band steadily decreases while
a new feature grows inside the bandgap of the bulk insulating STO.
After 1 hour, at $t_{f}$, the intensity of the valence band is about
half the initial value at $t_{0}$ ($I_{f}/I_{0}\sim0.5$). This change
coincides with the emergence and intensification of the signal at
$E_{F}$, which appears to be (meta)stable for hours under UHV conditions,
even when no beam is being applied (also noted in \cite{Meevasana2011}).

Observations similar to Fig.~\ref{fig:fig3}(a) prompted speculation
that photons generate oxygen vacancies that dope the surface \cite{Meevasana2011}.
Indeed, within the results obtained by us so far, photo-induced oxygen
vacancies and/or other defects remain as plausible hypotheses to explain
the origin of the carriers. However, by driving the decrease of the
valence band intensity much further than in previous STO studies,
the measurements here lead to new questions about the interpretation
of this particular behavior. For example, the universal FS in Figs.~\ref{fig:fig1}
and \ref{fig:fig2} corresponds to doping of only the $t_{2g}$ states
on the order of 0.1 $e^{-}$ per unit cell, whereas in principle 50\%
oxygen-vacant STO (nominally 3 $e^{-}$ per unit cell for SrTiO$_{1.5}$)
would be expected to have completely occupied $t_{2g}$ bands and
an FS composed of $e_{g}$ bands with 1 $e^{-}$ per unit cell. Consequently,
supposing that the O $2p$ intensity loss were purely due to oxygen
depletion from the surface of STO, one should conclude that something
like 90\% of the electrons localize. Alternatively, we can consider
that some other phenomenon (perhaps in addition to a limited amount
of oxygen loss) significantly contributes to the decrease in valence
band spectral weight.

To address this issue, Fig.~\ref{fig:fig3}(b) shows XPS of the O
$1s$ core level as a function of irradiation time. The spectra were
taken at nominal $t_{0}$ (established at a newly exposed spot) and
later, at $t_{f}$, after a dose of 47-eV photons approximately equivalent
to the 1 hour of irradiation in Fig.\textbf{~}\ref{fig:fig3}(a).
The photon energy ($h\nu=580$ eV for all core levels) was chosen
to closely match the kinetic energies of the O $1s$ photoelectrons
to those of the valence band states studied with $h\nu=47$ eV. As
a result, for the O $2p$ and O $1s$ peaks in Figs.~\ref{fig:fig3}(a)
and \ref{fig:fig3}(b), respectively, the photoelectron escape depths
are equal, and thus the probing regions of the two techniques are
identical. As the irradiation proceeds, the O $1s$ spectrum becomes
asymmetrically distorted by transferring weight to the high binding
energy sides of the peak. However, despite the change in the lineshape,
the total O $1s$ signal intensity remains largely stable under photon
irradiation. Integrating over the whole peak, the total countrate
intensity of O $1s$ at $t_{0}$ and $t_{f}$ is conserved to within
about 1\%, in stark contrast to the behavior of the oxygen valence
band in Fig.~\ref{fig:fig3}(a). The Ti $2p$ and Sr $3d$ core levels,
shown in Figs.~\ref{fig:fig3}(c) and \ref{fig:fig3}(d), also undergo
changes in their lineshapes, with the Ti $2p$ peaks in particular
showing a significant redistribution ($\sim50$\%) from Ti$^{4+}$
to Ti$^{3+}$ states. Like O $1s$, however, the energy-integrated
intensities of these core levels are conserved to within a few percent.
Various effects could account for the lineshape changes. For instance,
asymmetric skewing of the O $1s$ and Sr $3d$ peaks toward deeper
binding energy may be related to photohole screening in the metallic
state \cite{Doniach1970} and/or certain chemical changes, such as
the possible formation of surface SrO \cite{Chambers2012,SupMat}. 

The different behaviors of the valence band and core level signals
as a result of irradiation should be understood in terms of the fundamentally
different states being probed. Core level electrons are localized
with well-defined orbital characters. By contrast, the orbital characters
of valence states (and hence their photoemission matrix elements \cite{Damascelli2004})
may change, and/or such electrons may spatially redistribute in the
surface region, thus altering their visibility in photoemission ---
even absent a change in the surface composition \cite{SupMat}. Hence
the disproportionality between the changes in the valence band and
XPS intensities under irradiation is a signature of non-negligible
orbital/spatial changes of the near-$E_{F}$ states during the formation
of the surface metal.

Finally we note that during the beam exposure, the FS quickly saturates
to a relatively steady volume, while the signal intensity at the Fermi
level continues to grow. This is demonstrated in Fig.~\ref{fig:fig3}(e),
which compares total counts near the Fermi level (integrated within
a single $E$-vs.-$k_{x}$ slice through $k_{y}=0$ from -200 meV
up to $E_{F}$) on the left axis with the total FS volume of the $d_{xy}$
rings on the right axis as a function of time. The result indicates
that impinging photons do not significantly influence the carrier
concentration beyond a certain limit; they merely activate an increasingly
large area of the sample surface to become metallic at a uniformly
fixed carrier density, thus brightening the signal seen at the Fermi
level. Moreover, as the signal inensifies, the ARPES features appear
to sharpen while remaining $1\times1$ ordered in-plane, thus suggesting
that severe surface degradation does not occur \cite{SupMat}. This
behavior, considered alongside the universality of the fully formed
surface state with respect to various sample preparations (Fig.~\ref{fig:fig2}),
self-consistently indicates that STO's surface transitions between
two stable configurations --- one non-metallic and the other having
a fixed density of free carriers with universal dispersions and distinct
dimensionalities.

Despite clarifying the electronic structure of STO's surface, important
questions surround the origin of the carriers and the microscopic
process leading to the formation of the metallic state. For instance,
various defects such as O vacancies or excess Sr might dope the surface,
and even small amounts of defects allowed within the XPS presented
so far (i.e., the roughly 1\% reduction in O $1s$ intensity) could
be sufficient to explain the observed FS volume. However, it is surprising
to find the same electronic structure and surface carrier density
over such a broad range of sample preparations. This includes samples
annealed \emph{in situ} starting from predominantly TiO$_{2}$-terminated
wafers as in Fig.~\ref{fig:fig2}(a)--(d), as well as cleaved surfaces
of various annealed samples \cite{Santander-Syro2011} where the nature
and concentration of defects are likely to be significantly different
\cite{Guisinger2009}. Furthermore, as discussed, irradiating STO
has a profound effect on the electrons' initial states that goes beyond
merely doping the system in a rigid band manner. It is natural to
think this corresponds with a widespread structural change in the
surface region that is triggered directly by photons \cite{Nozawa2005,Qiu2005}
and/or indirectly by relatively dilute photo-induced defects. Along
these lines, one can propose that some common structural element of
the surface conducting state (e.g., intra-layer polar buckling as
found in STO-based interface metallic systems \cite{Pauli2011,Cantoni2012,Salluzzo2013}
and even bare STO surfaces \cite{Bickel1989,Hikita1993,Ikeda1999})
may be an important link between the variously treated samples that
helps to explain the universality of their surfaces' electronic properties,
despite nominally different compositions. Thus obtaining a full understanding
of the origins of the photo-induced spectral changes and their relation
to the surface metallic state is a pressing matter that should prompt
further investigations.

In conclusion, we have shown that the metallic state in the surface
region of SrTiO$_{3}$ is composed of two kinds of confined carriers
occupying quasi-2D $d_{xy}$ and energetically lifted nonbulklike
3D $d_{xz}$/$d_{yz}$ bands. Moreover we find evidence that a process
of generating metallicity at the surface of STO by photon irradiation
involves a substantial change in the initial states of the valence
electrons. Once formed, the metallic surface band structure and carrier
density of both types of electrons are essentially universal with
respect to diverse preparations of the samples. Similar electronic
structure is likely to be relevant in confined surface/interface conducting
states of related oxide systems. One example is KTaO$_{3}$ \cite{King2012,Santander-Syro2012},
whose metallic surface bands qualitatively resemble STO. There are
also similarities to conducting STO-based interfaces, where there
is evidence for two types of carriers and splitting of the $d_{xy}$
and $d_{xz}$/$d_{yz}$ states \cite{Zhou2011,Salluzzo2013,Chang2013a}.
Furthermore, in LaAlO$_{3}$/STO it is predicted that the $d_{xy}$
and $d_{xz}$/$d_{yz}$ states should spatially segregate along the
$z$-axis in a manner qualitatively consistent with the dimensionalities
of the respective FS components seen here on bare STO \cite{Delugas2011,Zhong2013a}.
Thus these new details of the electronic structure of STO's surface
state should be valuable for understanding, creating, and manipulating
functional oxide surfaces and interfaces.
\begin{acknowledgments}
{\small{}Experiments were conducted at the Surface/Interface Spectroscopy
(SIS) beamline of the Swiss Light Source within the Paul Scherrer
Institut in Villigen, Switzerland. We are grateful for valuable conversations
with J.~H.~Dil, V.~N.~Strocov, M.~Kobayashi, C.~Quitmann, A.~Uldry,
A.~F.~Santander-Syro, F.~Fortuna, E.~Rotenberg, R.~Claessen,
and F.~Miletto Granozio. M.~M.~was partly supported by the Swedish
Foundation BLANCEFLOR Boncompagni-Ludovisi n\'{e}e Bildt. M.~F.~acknowledges
financial support from the Swiss National Science Foundation (Project-No
PMPDP2\_128995). J.~B., H.~E., and J.~Min\'{a}r acknowledge financial
support from the Deutsche Forschungsgemeinschaft (FOR 1346), the Bundesministerium
f\"{u}r Bildung und Forschung (05K13WMA), and CENTEM (CZ.1.05/2.1.00/03.0088).
C.~M.~was partially supported by the Swiss National Science Foundation
and its NCCR MaNEP.}{\small \par}

\medskip{}

\end{acknowledgments}

\inputencoding{latin1}$^{*}$nicholas.plumb@psi.ch

$^{\dagger}$Current address: \inputencoding{latin9}\foreignlanguage{english}{Diamond
Light Source, Harwell Science and Innovation Campus, Didcot, Oxon,
OX11 0DE, United Kingdom}

\selectlanguage{english}%
$^{\ddagger}$milan.radovic@psi.ch

\putbib
\end{bibunit}

\pagebreak{}

\begin{bibunit}
\setcounter{figure}{0}
\renewcommand{\thefigure}{SM.\arabic{figure}}

\part*{Supplemental Material for ``Mixed Dimensionality of Confined Conducting
Electrons in the Surface Region of SrTiO$_{3}$''}

\section*{Supplemental Methods}

\subsection*{Samples}

SrTiO$_{3}$ and Nb-SrTiO$_{3}$ samples measuring approximately $5\times10\times0.5$
mm$^{3}$ were cut and polished on the (001) surface by the sample
provider (SurfaceNet GmbH). As a first step, the highly TiO$_{2}$-terminated
surfaces were prepared by us using a buffered HF etching and oxygen
annealing process similar to treatments described elsewhere \cite{Koster1998,Ohnishi2004}.
The details are as follows: We soaked the samples in deionized water
(30 min) and then transferred the wet wafers to a buffered HF solution
for a short period of etching (30 s). To immediately terminate the
etching, we then rapidly washed the samples in a sequence of deionized
water baths before drying them. Following etching, we then annealed
the samples at 1000 \textdegree C in flowing high-purity oxygen at
ambient pressure for about 2 h. Subsequently the samples underwent
\emph{in situ} treatments, beginning with annealing in 100 mbar O$_{2}$at
550 $^{\circ}$C for in order to clean the surfaces and obtain nominally
fully oxygenated samples \cite{Nishimura1999}. Some samples then
underwent additional \emph{in situ} annealing in UHV in order to create
oxygen vacancies or other defects. Each sample's final step of \emph{in
situ} treatment is summarized in Table \ref{tab:tab1}. 

\begin{table}[bh]
\protect\caption{\label{tab:tab1}\emph{In situ} treatments for the samples in Fig.~2(a)--(d)
of the main text.}
\begin{ruledtabular}

\begin{tabular}{cccccc}
Fig.~2 & Sample & Environment & $T$ ($^{\circ}$C) & $t$ (h) & Color\tabularnewline
\hline 
(a) & STO & O$_{2}$, 100 mbar & 550 & 2 & Clear\tabularnewline
(b) & STO & $\sim10^{-9}$ mbar & 300 & 15 & Clear\tabularnewline
(c) & STO & $\sim10^{-9}$ mbar & 720 & 1 & Black\tabularnewline
(d) & Nb-STO & O$_{2}$, 100 mbar & 550 & 2 & Black\tabularnewline
\end{tabular}

\end{ruledtabular}
\end{table}

\begin{figure}
\includegraphics[width=1\columnwidth]{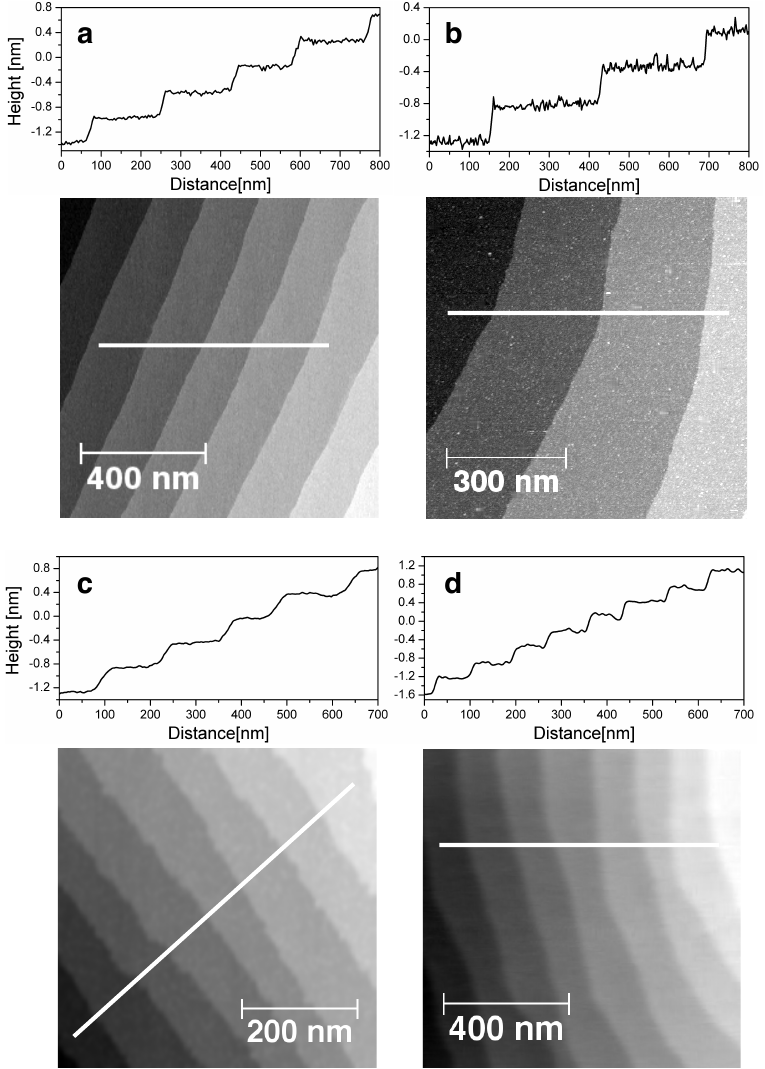}

\protect\caption{\textbf{\label{fig:figSI1}}Corresponding AFM and STM data of samples
in Fig.~2(a)--(d) of the main text (also summarized in Table \ref{tab:tab1}).
(a) Non-contact atomic force microscopy (AFM) on STO annealed in O$_{2}$.
(b) STM on STO annealed at 300 \textdegree C in UHV (1.0 nA of tunneling
current and bias voltage of 1.0 V). (c) STM on STO sample annealed
at 720 \textdegree C in UHV (tunneling current 0.5 nA, bias voltage
1.5 V). (d) Non-contact AFM on Nb-STO. The line profiles above each
image show that the terraces seen on the samples have unit cell step
height (0.39 nm), indicating they are predominantly single-terminated.}
\end{figure}

\begin{figure}
\includegraphics[width=1\columnwidth]{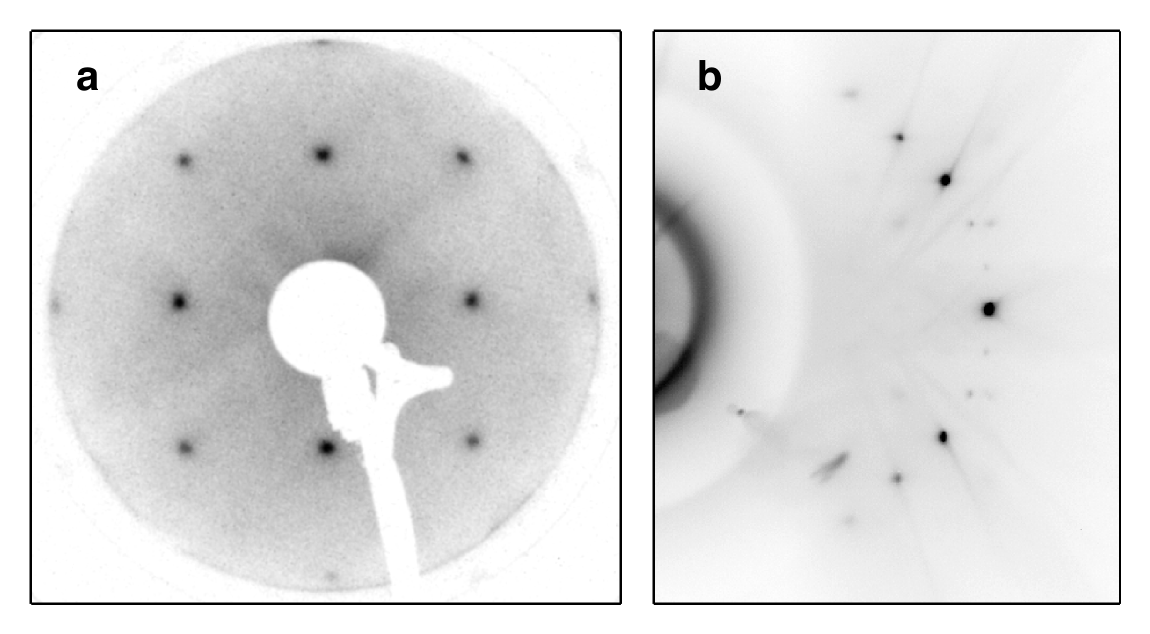}

\protect\caption{\textbf{\label{fig:figSI2}}Electron diffraction images from STO.
(a) Typical unreconstructed $1\times1$ LEED pattern, in this case
obtained from the Nb-STO sample (Fig.~1 and Fig.~2d of the main
text). The incident electron kinetic energy was 91 eV. (b) Typical
RHEED pattern from STO.}
\end{figure}

The high-quality surfaces were verified by scanning probe microscopy
performed \emph{after} the samples were\emph{ in situ} annealed and
studied by ARPES, which showed terraces separated by single-unit-cell
steps of $a=3.9$ \AA{} (Fig.~\ref{fig:figSI1}). Low-energy electron
diffraction (LEED) revealed that the surfaces are $1\times1$ ordered
(Fig.~\ref{fig:figSI2}a), with the possible exception of extremely
weak peaks corresponding to (twinned) $2\times1$ ordering that could
be seen on the sample annealed at 720 \textdegree C (featured in Fig.~2c
of the main text). Reflection high-energy electron diffraction (RHEED)
patterns additionally demonstrated the high crystallinity of the surfaces
(Fig.~\ref{fig:figSI2}b). We have indexed the Kikuchi lines and
find that the pattern is in excellent agreement with $1\times1$ surface
structure. Following the annealing procedures detailed in the main
text, the samples were not only visually different, but also exhibited
different conducting properties once removed from the vacuum chamber.
The resistances of the O$_{2}$- (Fig. 2a) and low-temperature vacuum-annealed
(Fig. 2b) samples were too high to be measured by our probe, while
the high-temperature vacuum-annealed (Fig. 2c) and Nb-STO (Fig. 2d)
samples had resistances of 0.5 k$\Omega$ and 2 k$\Omega$, respectively,
at 250 K under fixed measurement geometry. 

\subsection*{Photoemission Spectroscopy}

Angle-resolved photoemission spectroscopy (ARPES) \cite{Damascelli2004}
and core level x-ray photoemission spectroscopy (XPS) measurements
were carried out at the Surface/Interface Spectroscopy (SIS) beamline
at the Swiss Light Source. The endstation features a six-axis liquid
helium cryostat CARVING\texttrademark{} manipulator and a VG Scienta
R4000 hemispherical electron analyzer. The typical combined energy
resolution of the beamline and analyzer during our measurements was
about 20 meV. The pressure inside the UHV analysis chamber is on the
order of $10^{-11}$ mbar. The experiments were carried out at low
temperature (10--15 K). ARPES measurements of the valence and conduction
bands were performed using circular polarized light at photon energies
between 35 and 95 eV. XPS measurements were done using linear $p$-polarized
light at $h\nu=580$ eV with the sample oriented near normal emission.

\subsubsection*{ARPES: 3D $k$-space mapping}

In order to compile the 3D Fermi surface featured in Fig.~1, we varied
the photon energy from 35 to 95 eV in 2-eV steps while collecting
2D $k_{x}\text{--}k_{y}$ maps in the second Brillouin zone centered
about $k_{x}=2\text{ }\pi/a,k_{y}=0$. The momentum along (001), $k_{z}$,
was calculated using the free electron final state approximation \cite{Damascelli2004}:
\[
k_{z}=\frac{1}{\hbar}\sqrt{2m_{e}(E_{i}+h\nu-\Phi)\cos\theta+V_{0}},
\]
where $m_{e}$ is the free electron mass, $E_{i}$ is the initial
state relative to $E_{F}$, $h\nu$ is the photon energy, $\theta$
is the emission angle, $\Phi$ is the sample work function ($\approx4.5$
eV), and $V_{0}$ is the so-called ``inner potential''. The $k_{z}$
transformation was calculated using $V_{0}=14.5$ eV. This value gives
good results by eye, with no need to offset the resulting Fermi surface
along $k_{z}$. Moreover, this value of $V_{0}$ is in line with that
of other perovskite compounds, as well as crude expectations based
on the depth of the oxygen valence band minimum. As a final step,
the data was symmetrized about $k_{x}=2\text{ \ensuremath{\pi}/a}$
to minimize matrix element effects, leading to the map presented in
Fig.~1e. No normalization was applied to the data. 

Finally, concerning the details of the crystal structure, it should
be noted that throughout the text, we regard the crystal as cubic
($a=b=c$). In reality, bulk STO is slightly tetragonal at low temperature,
which may arise in conjunction with the surface rumpling \cite{Singh2012},
but the change in unit cell dimensions ($c/a=1.00056$) is too small,
e.g., to have a noticeable effect on the $k$ units conversion for
ARPES.

\subsubsection*{Dispersion analysis: Tight-binding fits}

Several useful parameters can be extracted from the ARPES data for
direct comparison with other probes, such as transport measurements.
These include the area/volume of each FS component (corresponding
to the carrier electron density) as well as each band\textquoteright s
effective mass $m^{*}=\hbar^{2}\left[\nabla_{\boldsymbol{k}}^{2}\epsilon(\boldsymbol{k})\right]^{-1}$
evaluated at the Fermi momentum. Estimates of these parameters based
on tight-binding fits of the data are given in Table \ref{table:tableSI1}.
All fits and extracted values assume double spin occupancy of the
bands (i.e., that the bands are non-spin-polarized). For the ellipsoids,
$m^{*}$ varies as a function of position on the Fermi surface. Values
are quoted along the \textquotedblleft light\textquotedblright{} and
\textquotedblleft heavy\textquotedblright{} axes in the $x$-$y$
plane, as well as the $z$ axis. The final column notes the energy
of the dispersion relative to $E_{F}$ at the bottom of each band.
For the tight-binding fits used in determining the effective mass
values, it was found that the measured band dispersions could be modeled
well by a simple equation \cite{Popovic2008} adapted to have adjustable
parameters in three dimensions: where $a$ is the lattice constant
(3.9 \AA). For each band we solved the system of equations for $\epsilon(0)=\mu-E(0)$
and $\epsilon(k_{F})=\mu$, where $E(0)$ is the measured band bottom,
$\mu$ is the chemical potential, and the Fermi momentum $k_{F}$
is evaluated along each of the principal axes {[}i.e., $(k_{F},0,0)$,
$(0,k_{F},0)$, and, for 3D bands, $(0,0,k_{F})${]}. The extracted
model parameters are as follows: $3d_{xy}$ \textquotedblleft outer
ring\textquotedblright{} \textrightarrow{} $V_{x}=V_{y}=-0.920$ eV,
$V_{z}=0$, $\mu=-1.611$ eV; $3d_{xy}$ \textquotedblleft inner ring\textquotedblright{}
\textrightarrow{} $V_{x}=V_{y}=-1.009$ eV, $V_{z}=0$, $\mu=-1.908$
eV; $3d_{xz}$ ($3d_{yz}$) ellipsoid \textrightarrow{} $V_{x}=-0.076$
eV (-2.076 eV), $V_{y}=-2.076$ eV (-0.076 eV), $V_{z}=-0.083$ eV,
$\mu=-2.185$ eV. This parameterization regards the $3d_{xy}$ inner
and outer ring Fermi surface components as purely 2D, while the $3d_{xz}$/$3d_{yz}$
ellipsoids are treated as 3D. This is an \emph{ad hoc} simplification
implemented for the purpose of estimating the effective mass and carrier
densities of each of the Fermi surface components. Naturally, however,
the dimensionality is somewhat ill-defined in the 2D-3D crossover
regime. 

\begin{table*}
\caption{\label{table:tableSI1}\textbf{Estimates of the electron density for each Fermi surface component.} The approximate effective masses, determined from tight-binding model fits, are also given. The fitting and values assume double spin occupancy of each band.}
\begin{ruledtabular}

\begin{tabular}{cccc}
Component & Electron density & $m^{*}$($m_{e}$) & Band bottom rel. to $E_{F}$ (meV)\tabularnewline
\hline 
$3d_{xy}$ ``outer ring'' & 0.084 $\text{e}^{-}/a^{2}=5.5\times10^{13}$ cm$^{-2}$ & 0.7 & -230\tabularnewline
$3d_{xy}$ ``inner ring'' & 0.036 $\text{e}^{-}/a^{2}=2.4\times10^{13}$ cm$^{-2}$ & 0.6 & -110\tabularnewline
 &  & 0.4 ($x$-$y$ light axis) & \tabularnewline
$3d_{xz}$/$3d_{yz}$ ellipsoids & 0.01 $\text{e}^{-}/a^{3}=2\times10^{20}$ cm$^{-3}$ (each) & 19 ($x$-$y$ heavy axis) & -50\tabularnewline
 &  & $\sim15$ ($z$ axis) & \tabularnewline
\end{tabular}\end{ruledtabular}
\end{table*}

\subsubsection*{XPS: Core level spectral weight analysis}

To analyze the XPS data, we removed an \textquotedblleft extrinsic\textquotedblright{}
background {[}i.e., electrons not originating from the feature(s)
of interest{]} from each spectrum and then summed the total spectral
intensity over each relevant energy range. For the analysis of the
Ti $2p$ and Sr $3d$ peaks, it was sufficient to remove simple offset
backgrounds determined by the mean value of the flat signal on the
low-binding-energy side of the spectra, away from the peaks. The backgrounds
of the O $1s$ peaks had more complicated, non-negligible energy dependences.
To determine and remove these backgrounds, we fit each O $1s$ spectrum
with an exponential function over the binding energy range of 517
to 547 eV, ignoring (i.e., masking) the window from 528 to 536 eV. 

\begin{figure}
\includegraphics[width=0.8\columnwidth]{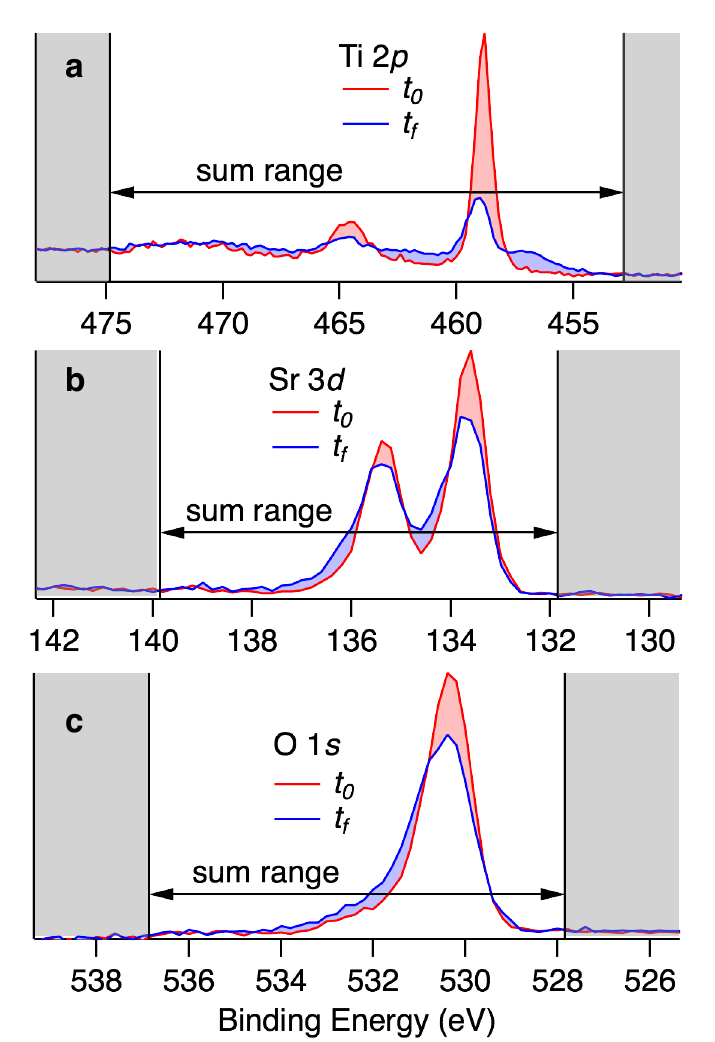}

\protect\caption{\label{fig:figSI3}Summing ranges of the quantitative XPS analysis.
(a)--(c) Ti $2p$, Sr $3d,$and O $1s$ core level spectra at initial
time $t_{0}$ and final time $t_{f}$.}
\end{figure}

Our analysis is concerned with all electrons associated with each
core-level feature, including inelastically scattered electrons and
shake-up/-off structures. Thus, for example, the summing range applied
to the Ti $2p$ spectra (Fig.~\ref{fig:figSI3}a) includes the broad
shake-up hump located at a binding energy of about 471 eV, and any
inelastically scattered electrons that may give the spectrum a steplike
offset (i.e., a Shirley-like background \cite{Shirley1972}) are counted
in the total spectral weight as well. Other related spectral features
are present at even higher binding energy. Such features overlap well
between the $t_{0}$ and $t_{f}$ spectra, so that integrating the
spectral weight out to even higher binding energy brings the spectral
weight conservation ratio, $I_{f}/I_{0}$, closer to unity. For this
reason, we report $I_{f}/I_{0}>0.94$ for Ti $2p$. Details of the
summing ranges for all peaks, are shown in Fig.~\ref{fig:figSI3}.
We have verified for all spectra that any reasonable changes to the
background subtraction procedures and signal integration windows have
no substantive effect on the overall outcome, meaning that the total
intensity from each core level spectrum is still conserved to within
a few percent. This stands in contrast to the spectral weight transfers
from the oxygen valence band to the in-gap and Fermi level states
and from the Ti$^{4+}$ to the Ti$^{3+}$ core level structures, wherein
those features shrink or grow by roughly a factor of two. Thus the
particulars of the data analysis are presented here for completeness
but ultimately have little bearing on the key findings.

\subsubsection*{Near-Surface Ti$^{3+}$: Emission Angle Dependence of Ti $2p$ XPS }

\begin{figure}
\includegraphics[width=0.8\columnwidth]{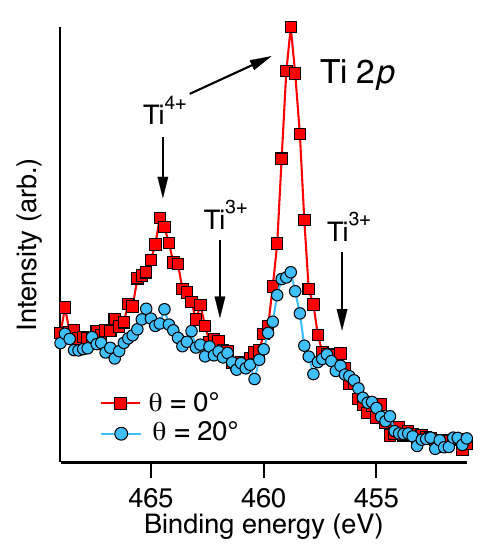}

\protect\caption{\label{fig:figSI4}Emission angle ($\theta$) dependence of the Ti
$2p$ spectrum illustrating that Ti$^{3+}$ states are concentrated
near the surface.}
\end{figure}

We performed XPS as a function of the emission angle $\theta$ in
order to further verify that the Ti$^{3+}$ states associated with
the observed metallic state are concentrated near the surface. Photoemission
is a surface sensitive technique due to the low mean free path $\lambda$
of the electrons in the solid, which is typically considered to be
$\lesssim1$ nm in the kinetic energy range employed here \cite{Seah1979}.
The contribution to the total XPS signal from individual atoms follows
a Beer\textquoteright s Law relation with respect to the escape depth
that the emitted electrons experience. Hence the signal intensity
$I(z)$ from an atom at depth $z$ from the surface is
\[
I(z)=I(0)\exp\left(-\frac{z}{\lambda\cos\theta}\right)
\]
where $\theta=0$ is normal to the sample surface. Figure \ref{fig:figSI4}
shows Ti $2p$ spectra at $\theta=0^{\circ}$ and $\theta=20^{\circ}$
obtained from a Nb-STO sample using $h\nu=600$ eV. The Ti$^{4+}$
peaks shrink at higher emission angles, while the Ti$^{3+}$ features
are roughly constant. This is consistent with Ti$^{3+}$ being concentrated
near the sample surface where the XPS intensity will have little-to-no
angular dependence, while the Ti$^{4+}$ states are buried deeper
and thus very sensitive to the angle. Aware of the fact that the Ti$^{3+}$
concentration may change during the measurement due to the irradiation
effects described in the main text, we took care to perform the measurements
quickly, at relatively low beam intensity, and while maintaining the
same position of the beam on the sample. Moreover, we performed the
measurement at $\theta=20^{\circ}$ first, followed by the one at
$\theta=0^{\circ}$. If anything, the photo-assisted changes would
cause the Ti$^{4+}$ XPS intensity to decrease relative to Ti$^{3+}$
between these measurements. Instead, Ti$^{4+}$ increased relative
to Ti$^{3+}$, so the change can safely be attributed to the difference
in emission angle. 

\subsubsection*{Sample grounding}

\begin{figure}
\includegraphics[width=0.7\columnwidth]{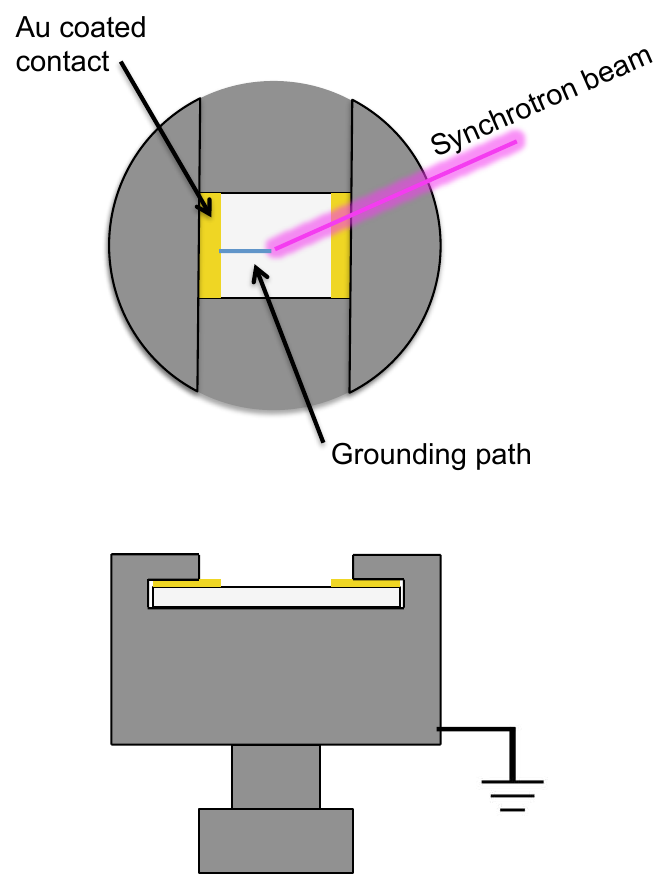}\protect\caption{\label{fig:figSI5}Illustrations of the grounding scheme for bulk
insulating STO samples. By scanning the sample under the synchrotron
beam, a long-lived conducting pathway can be traced on the surface
of an otherwise insulating sample by the process described in the
main text. The path can be connected to gold deposited on the sample
surface, which improves contact to the UHV transferable sample holder.
The top and bottom panels are top and side views of the holder, respectively.}
\end{figure}

Special care was taken to alleviate charging during the ARPES measurements.
After the initial etching/annealing process described above, we evaporated
gold coatings near the edges of the samples. (A large central area
of each sample was protected by a mask during deposition.) Each finished
wafer with gold-coated edges was then fixed by clamps into a UHV transferable
sample holder. The metal clamps made direct contact with the gold
coatings on the STO. Once transferred to the ARPES system, each sample
was further prepared \emph{in situ} by one of the procedures described
in Table 1 of the main text. Samples with insulating bulks, such as
those in Fig. 2(a) and 2(b) of the main text, have a natural tendency
to charge during ARPES, which can be problematic for the measurement.
To overcome this issue, we made use of the long-lived photoinduced
conductivity that can form on STO. By slowly scanning the sample under
the synchrotron beam, we could trace conduting pathways from a gold
contact to any arbitrary point on a sample (see Fig.~\ref{fig:figSI5}). 

\section*{Supplemental Background}

\subsection*{Initial State Effects in ARPES}

ARPES measures the single-particle removal function, $F(\boldsymbol{k},E)$,
multiplied by the photoexcitation matrix elements $M_{fi}=|\langle\phi_{\boldsymbol{k},f}|\boldsymbol{A}\cdot\boldsymbol{p}|\phi_{\boldsymbol{k},i}\rangle|^{2}$
(see, e.g., \cite{Damascelli2004}). $F(\boldsymbol{k},E)$ reflects
the band structure, taking into account the quasiparticle lifetime
due to the many-body self-energy. $M_{fi}$ is the dipole transition
between the initial and final $\boldsymbol{k}$ states, $|\phi_{\boldsymbol{k},i}\rangle$
and $\langle\phi_{\boldsymbol{k},f}|$, respectively, which depends
on the incoming photon polarization and energy within the vector potential
$\boldsymbol{A}$. The $\boldsymbol{k}$-dependent cross section is
then $\sigma=F(\boldsymbol{k},E)M_{fi}$. 

Although $F(\boldsymbol{k},E)$ and $M_{fi}$ correspond with different
phenomena \textendash{} one the quasiparticle dispersion and the other
the radiative transition \textendash{} they are coupled through the
states, in the sense that if the nature of $\left|\phi_{\boldsymbol{k},i}\right\rangle $
and/or $\left\langle \phi_{\boldsymbol{k},f}\right|$ changes in a
way that alters $M_{fi}$, this should be reflected in the dispersion
information of $F(\boldsymbol{k},E)$ as well. Thus, there is no loss
of generality to associate changes in valence band spectral weight
via the cross section with changes in the initial states. In principle,
the final states can be included in this statement as well, but this
appears to be ruled out as a major effect in the present work, since
the O $1s$ XPS data shows near conservation of total spectral weight
after irradiation, even though the kinetic energy (i.e., final state)
is matched to that of the valence band probed with $h\nu=47$ eV,
where the $\sim50\%$ decrease is observed.

A second possible effect, also mentioned in the main text, could be
related to the escape depth, $\lambda(E_{\text{kin}})$. Supposing
that surface oxygen valence electrons redistribute into the subsurface
by some charge transfer mechanism, those states would become exponentially
less visible to the photoemission process as a function of their depth,
leading again to a discrepancy between the intensity behavior of the
valence band and O $1s$ core level. This effect would occur in addition
to matrix element-related intensity changes resulting from the charge
transfer.

\subsection*{Changes in the XPS Lineshapes }

As shown in Fig.~3 of the main text, irradiating the sample and thereby
forming the surface metal induces an asymmetric redistribution of
the spectral intensity of the XPS peaks toward deeper binding energy.
This is particularly evident in the O $1s$ and Sr $3d$ peaks. Such
changes could be qualitatively consistent with well-known screening
effects first theoretically explained by Doniach and \v{S}unji\'{c}
\cite{Doniach1970}. However, alternate, or additional, factors could
be at play, including certain changes in surface chemistry. Crucially,
the results reported in our study put significant constraints on the
types and mechanisms of chemical changes that may be considered, and
in what ways these changes could potentially be relevant to the formation
of the surface metallicity. Namely, the near-conservation of energy-integrated
XPS weight implies that such changes would necessarily be subtle,
involving very little change in the concentration of each atomic species
at the surface. Moreover, since the carrier density and other electronic
properties of STO\textquoteright s metallic surface are highly robust
with respect to diverse sample preparations (Fig.~2 of the main text),
including variously prepared cleaved samples employed in other studies
\cite{Santander-Syro2011,Meevasana2011}, any possible relevance to
the surface conductivity appears to be indirect in the sense that
they are not proportional to the surface carrier density.

To take a specific example, we consider the evolution of the shape
of the Sr $3d$ and O $1s$ peaks, which raises a question of whether
Sr or SrO may form at the surface as a consequence of irradiation.
Certain mechanisms to explain such a change are not consistent with
the observations. Supposing, for instance, that Sr were to migrate
from the subsurface to the surface, the Sr signal should intensify
due to significantly reduced scattering. (The inelastic mean free
path in our experiments is on the order of a single unit cell.) In
contrast, experimentally the total Sr intensity is nearly conserved
over the course of irradiation, perhaps with a slight decrease of
about 1\%. 

It is more plausible to instead suppose that a certain amount of Sr
preexists on the otherwise predominantly TiO$_{2}$ surface, and the
effect of irradiation is to cause oxygen to rearrange forming SrO
(leaving behind surface Ti$^{3+}$), thus leading to the observed
lineshape changes. This mechanism has the appeal of maintaining the
same concentrations of Ti, Sr, and O at the surface, consistent with
the findings. It also offers a plausible explanation of the Sr and
O lineshapes, as well as the appearance of Ti$^{3+}$ shoulders adjacent
to the main Ti$^{4+}$ peaks. However, as demonstrated in our experiments,
the observed surface carrier concentration is fixed at a relatively
low value on the order of about 0.1 $e^{-}/a^{2}$ for diversely prepared
samples, including the cleaved samples of previous ARPES studies,
where one expects the surface concentration of Sr be substantively
different from the treated samples in the present work. In addtion,
there is no evidence from ARPES, LEED, or RHEED to connect the metallicity
with an in-plane atomic reconstruction that might account for the
high stability of this or any other surface carrier density. Our data
therefore imply that even if photoinduced changes in surface chemistry
might be relevant to the metallicity, they are connected by an indirect
mechanism. In other words, perhaps they act to dope the system, but
if so, their actual contribution to the carrier density is tightly
constrained by some different behavior that plays a dominant role
in governing the electronic surface properties. We speculate that
this could be related, e.g., to polar surface distortions, which have
been previously observed on STO \cite{Bickel1989,Hikita1993,Ikeda1999}
and might strongly influence the electrostatic conditions that confine
the carriers, but further studies are necessary to ellucidate the
mechanism behind the robust surface carrier density.

\subsection*{Further details regarding changes in the ARPES spectra of the metallic
bands during irradiation}

\begin{figure*}
\includegraphics[width=1\textwidth]{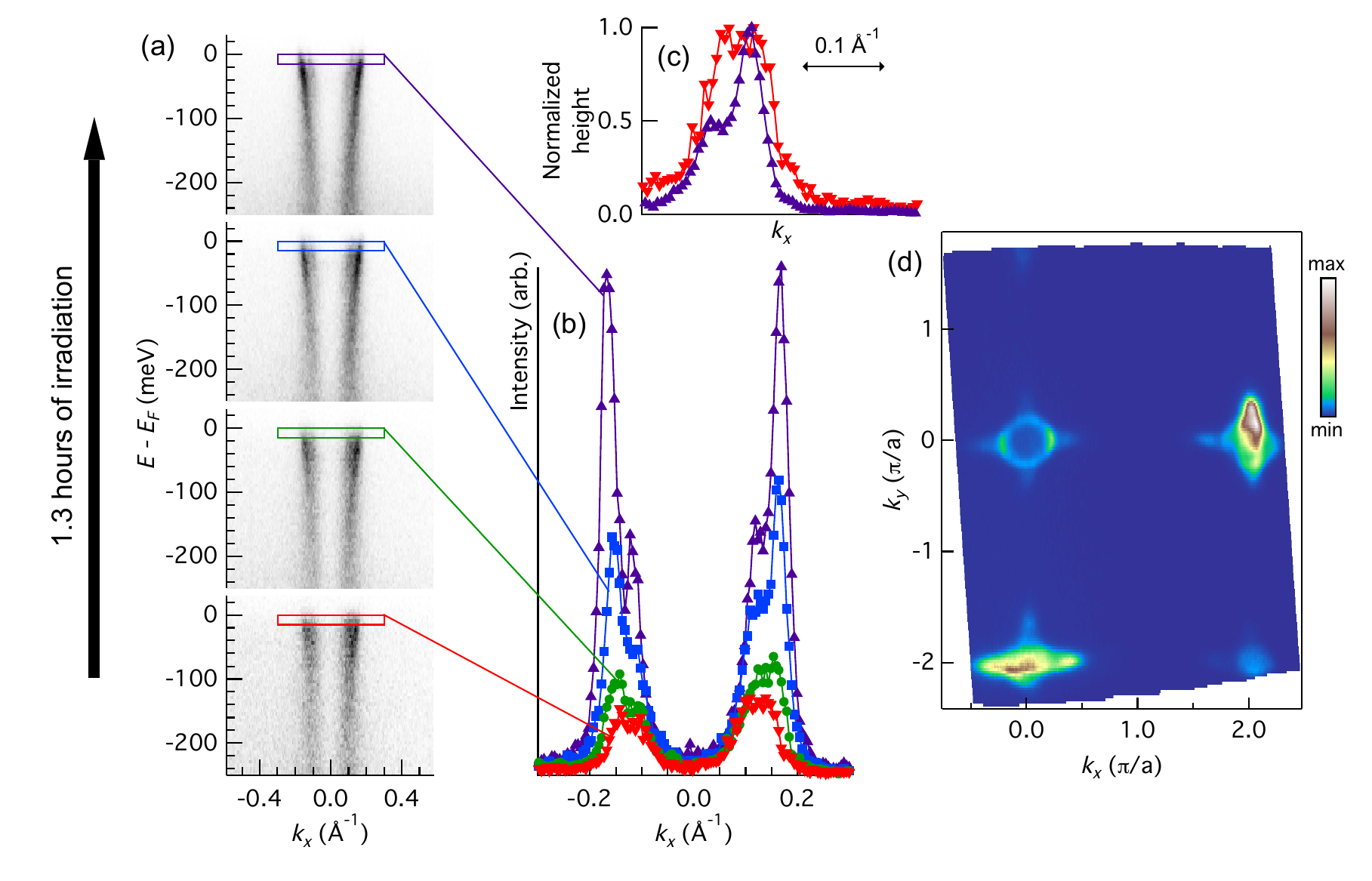}

\protect\caption{\label{fig:figSI6}(a) Dispersions of the $d_{xy}$ bands obtained
at different times from the same spot, over the course of 1.3 hours
of irradiation from the synchrotron beamline. The spectra were acquired
close to the $\Gamma$ point in the first Brillouin zone (i.e., centered
around $k_{x}=k_{y}=0$) using 47-eV circularly polarized photons.
The differences in the dispersions when compared with, e.g., Fig.~1(d)
of the main text are due to a matrix element effect that suppresses
the signal near $\boldsymbol{k}=0$, making the band bottoms invisible
in the first zone, although the $k_{F}$ points are the same. Note
that these are the same spectra from which the quantities in Fig.~3(e)
of the main text were extracted; this sample also corresponds to Fig.~
2(a) and the first line of Table \ref{tab:tab1}. (b) Corresponding
MDCs from near the Fermi level (integrated from -15 meV up to $E_{F}$,
shown by the boxed regions) for each spectrum in (a). (c) Comparison
of the first (red down triangles) and last (purple up triangles) MDCs
with their heights normalized to 1 in order to demonstrate the sharpening
of the spectral features during the irradiation, with the separation
of inner and outer $d_{xy}$ bands becoming clearer. (d) Fermi surface
map in the $k_{x}$-$k_{y}$ plane measured over multiple Brillouin
zones using 85-eV circularly polarized photons. The sample was Nb-STO
prepared in the same manner as for Fig.~2(d) and line 4 of Table
\ref{tab:tab1}. The lack of folded/replica Fermi surfaces shows that
the metallic state is $1\times1$ ordered in-plane. The temperature
was 11--15 K for all spectra.}
\end{figure*}

Given the extent of the spectral changes to the O $2p$ valence band,
as well as the lineshapes of the O $1s$, Sr $3d$, and Ti $2p$ core
levels, one may ask whether some large-scale chemical changes occur
during irradiation that either break down/degrade or reform the surface
while otherwise mostly maintaining the overall surface stoichiometry,
consistent with the (near) conservation of energy-integrated core
level spectral weight. (An example might be disproportionation of
the surface into SrO$_{x}$ and Ti$_{2}$O$_{3}$.) We point out that
if the surface were to degrade or undergo a reconstruction, one would
expect this to cause broadening of the dispersive ARPES features,
the appearance of replica (folded) Fermi surfaces, or both. No such
behavior is observed during irradiation. On the contrary, as shown
in Fig.~\ref{fig:figSI6}(a)--(c), the momentum distribution curves
(MDCs) of the metallic bands do not broaden, but instead \emph{sharpen},
as a result of irradiating the surface. Meanwhile the Fermi surface
has been measured over several Brillouin zones in the $k_{x}$-$k_{y}$
plane and has always been found to be $1\times1$ ordered {[}Fig.~\ref{fig:figSI6}(d){]}.

\putbib
\end{bibunit}
\end{document}